\documentclass[twocolumn,showpacs,preprintnumbers,amsmath,amssymb]{revtex4}

\usepackage{graphicx}
\usepackage{dcolumn}
\usepackage{bm}

\begin{document}

\preprint{}

\title{Super-additivity of quantum-correlating power}

\author{Xueyuan Hu}
\email{xyhu@iphy.ac.cn}
\author{Heng Fan}
\author{D. L. Zhou}
\author{Wu-Ming Liu}
\affiliation{Beijing National Laboratory for Condensed Matter
Physics, Institute of Physics, Chinese Academy of Sciences, Beijing
100190, China}
\date{\today}

\begin{abstract}
We investigate the super-additivity property of quantum-correlating power (QCP),
which is the power of generating quantum correlation by a local quantum channel.
We prove that, when two local quantum channels are used paralleled,
the QCP of the composed channel is no
less than the sum of QCP of the two channels. For local channels
with zero QCP, the super-activation of QCP is a fairly common
effect, and it is proven to exist widely except for the trivial case where both
of the channels are completely decohering channels or unitary
operators. For general quantum channels, we show that the
(not-so-common) additivity of QCP can be observed for the situation
where a measuring-and-preparing channel is used together with a
completely decohering channel.
\end{abstract}

\maketitle

\section{Introduction}
Contrary to quantum entanglement, which is a monotone under local
operations, general quantum correlations, such as quantum discord,
can be created and increased by local operations
\cite{PhysRevLett.107.170502,PhysRevA.85.032102,arXiv:1112.5700v1,PhysRevA.85.010102,PhysRevA.85.022108}.
This is understandable, as the local operations can turn some
classical correlation into quantum one. More and more evidences show
that quantum correlation is responsible for quantum information
processes, such as remote state preparation
\cite{arXiv:1203.1629v1,arXiv:1205.0251}, entanglement distribution
\cite{arXiv:1203.1264v2,arXiv:1203.1268v2}, quantum state
discrimination \cite{PhysRevLett.107.080401}, etc. The importance of
quantum correlation also lies in its close connection to quantum
entanglement
\cite{PhysRevLett.106.220403,PhysRevLett.106.160401,PhysRevLett.107.020502}.
The local creation of quantum correlation makes the preparation of a
quantum correlated state simple. Meanwhile, it provides a method to
study the properties of both the quantum correlations and local
quantum channels. Generally, a local channel is able to create quantum correlation if
and only if it is not a commutativity-preserving channel
\cite{PhysRevA.85.032102}. For qubit case, a
commutativity-preserving channel is either a completely decohering
channel or a unital channel
\cite{PhysRevLett.107.170502,PhysRevA.85.032102}, while for qutrit
case, the set of commutativity-preserving channels reduces to
completely decohering channels and isotropic channels
\cite{PhysRevA.85.032102}.

On solving the problem of whether a channel can create quantum
correlation, it is natural to investigate how much quantum
correlation can be created locally. For this purpose, QCP was
proposed as the maximum quantum correlation that can be created by a
given local quantum channel \cite{arXiv:1203.6149}. QCP not only
quantifies the amount of quantum correlation created by local
operation, but also serves as an inherent property of quantum
channels. It is of interest to investigate the effect caused by
using two channels together. We have given an example to indicate
the super-activation for QCP of two zero-QCP channels in Ref.
\cite{arXiv:1203.6149}. It implies that using two channels together
is more efficient in creating quantum correlations than using them
separately. It is straightforward to ask the following questions:
What kind of local channels has the property of super-activation of
QCP? Is super-additivity of QCP holds for general quantum channels?
We remark that additivity property is one of the most fundamental problems
in quantum information science,
for example, it is shown that communication capacity
using entangled inputs is superadditivity \cite{Hastings}.

In this paper, we prove that when the two channels have zero-QCP,
the super-activation can be observed except that the two channels
are both unitary operations or completely decohering channels. It
means that the super-activation of QCP is a common phenomenon. We
then prove that the answer to the first question is positive. When
one of the channels have positive QCP, there are still situations
where the QCP of the two channels are additive. The QCP of the
bi-channel which is composed of a measuring-and-preparing (MP)
channel and a completely decohering channel equals to that of the MP
channel. We consider the genuine quantum correlation to be
responsible for the super-additivity of QCP.

\section{Super-activation of quantum-correlating power}

Let us briefly recall the definition of QCP \cite{arXiv:1203.6149}
for local channels. The quantum-correlating power of a quantum
channel $\Lambda$ is defined as
\begin{equation}
\mathcal Q(\Lambda)=\max_{\rho\in\mathcal C_0}Q(\Lambda\otimes
I(\rho)).\label{QCP}
\end{equation}
Here $\mathcal C_0$ is a set of classical-quantum states, which can
be written as \cite{arXiv:0807.4490v1}
\begin{equation}
\mathcal C_0=\{\rho|\rho=\sum_iq_i\Pi_{\alpha_i}^A\otimes\rho_i^B\},
\end{equation}
and $Q$ is a measure of quantum correlation satisfying the following
three conditions: (a) $Q(\rho)=0$ iff $\rho\in\mathcal C_0$; (b)
$Q(U\rho U^{\dagger})=Q(\rho)$ where $U$ is a local unitary operator
on $A$ or $B$; (c) $Q(I\otimes\Lambda_B(\rho))\leq Q(\rho)$.

In Ref. \cite{arXiv:1203.6149}, we have given an example to show the
effect of super-activation for QCP of phase-damping qubit channels
$\Lambda^{\mathrm{PD}}$. Precisely, we consider the input state
$\rho=\frac14\sum_{i,j}|\psi_{ij}\rangle_{AA'}\langle
\psi_{ij}|\otimes|ij\rangle_{BB'}\langle ij|$ with
$|\psi_{00}\rangle=\frac{1}{\sqrt2}(|00\rangle+|11\rangle)$,
$|\psi_{11}\rangle=\frac{1}{\sqrt2}(|0+\rangle+|1-\rangle)$,
$|\psi_{01}\rangle=\frac{1}{\sqrt2}(|01\rangle-|10\rangle)$, and
$|\psi_{10}\rangle=\frac{1}{\sqrt2}(|0-\rangle-|1+\rangle)$. Here, the qubits $A$ and $A'$ belong to Alice while $B$ and $B'$ belongs to Bob. Hence, Alice and Bob are classically correlated initially. Now Let the
composed channel
$\Lambda^{\mathrm{PD}}_A\otimes\Lambda^{\mathrm{PD}}_{A'}$ acting on
the qubits $A$ and $A'$. The resulted state has nonzero quantum
correlation because
$[\Lambda^{\mathrm{PD}}\otimes\Lambda^{\mathrm{PD}}(\psi_{00}),
\Lambda^{\mathrm{PD}}\otimes\Lambda^{\mathrm{PD}}(\psi_{11})]
=\frac18\tilde
ip\sqrt{1-p}(\sigma^y\otimes\sigma^z+\sigma_z\otimes\sigma^y)\neq0$.
Therefore, the QCP of composed channel
$\Lambda^{\mathrm{PD}}\otimes\Lambda^{\mathrm{PD}}$ is nonzero.

Then what kind of zero-QCP channels has the property of
super-activation of QCP? Obviously, if the two channels are both
completely decohering channels, the composed channel is still a
completely decohering channel, which is not able to create quantum
correlation. Similarly, for the case that the two channels are both
unitary operators, the composed channel is a unitary operator too,
and thus has zero-QCP. In the following, we will prove that these
are the only two situations where super-activation of QCP is not
observed.

\emph{Theorem 1.} For two zero-QCP channels $\Lambda_1$ and
$\Lambda_2$, the QCP of the composite channel
$\Lambda_1\otimes\Lambda_2$ is non-zero except that both $\Lambda_1$
and $\Lambda_2$ are completely decohering channels or the unitary
channels.


\emph{Proof}. From Ref. \cite{PhysRevA.85.032102}, the QCP of
channel $\Lambda_1\otimes\Lambda_2$ is zero if and only if
$\Lambda_1\otimes\Lambda_2$ is a commutativity-preserving channel.
It means that, for any commutative two-particle states $\xi_1$ and
$\xi_2$, we have
\begin{equation}
[\Lambda_1^A\otimes\Lambda_2^{A'}(\xi_1),\Lambda_1^A\otimes\Lambda_2^{A'}(\xi_2)]=0.\label{criterion}
\end{equation}
Obviously, when both $\Lambda_1$ and $\Lambda_2$ are completely
decohering channels or unitary channels, the channel
$\Lambda_1\otimes\Lambda_2$ is also a commutativity-preserving
channel, Eq. (\ref{criterion}) holds. Otherwise, let $\Lambda_1$ not
be a completely decohering channel and $\Lambda_2$ not be a unitary
channel. Now we choose
$\xi_1=|0\rangle_{A}\langle0|\otimes|\psi\rangle_{A'}\langle\psi|$
and
$\xi_2=|\theta\rangle_{A}\langle\theta|\otimes|\psi^{\bot}\rangle_{A'}\langle\psi^{\bot}|$,
where $\langle\psi|\psi^{\bot}\rangle=0$ and $|\theta\rangle$ is an
arbitrary single-particle state. Then Eq. (\ref{criterion}) is
equivalent to
\begin{equation}
[\Lambda_1(|0\rangle\langle0|),\Lambda_1(\theta)]\otimes\Lambda_2(\psi)\Lambda_2(\psi^{\bot})=0.\label{criterion1}
\end{equation}
Here we label $\psi$ as the density matrix as the pure state
$|\psi\rangle$ and similar for $\psi^{\bot}$ and $\theta$. Since
channel $\Lambda_1$ is not a completely decohering channel, we can
always find a single-particle state $|\theta\rangle$ such that
$[\Lambda_1(|0\rangle\langle0|),\Lambda_1(\theta)]\neq0$. Meanwhile,
when $\Lambda_2$ is not a unitary channel, there exist two pure
orthogonal single-particle states $\psi$ and $\psi^{\bot}$ such that
$\Lambda_2(\psi)\Lambda_2(\psi^{\bot})\neq0$. Therefore, Eq.
(\ref{criterion1}) is violated. It completes the proof of Theorem 1.

In the above discussion, super-activation of QCP is due to the
non-classicality of particle $A$. An extreme example is that,
$\Lambda_1=I$ is an identity channel while $\Lambda_2$ is a
completely depolarizing channel, which is equivalent to the
``trace-out'' operation. Now we start with the state
$\rho_{AB}\otimes|0\rangle_{A'}\langle0|$, where
$\rho_{AB}\in\mathcal C_0$. After a two-particle unitary operator
$U_{AA'}$ on $A$ and $A'$, which does not create quantum correlation
on the left, the channel $I\otimes\Lambda_2$ is applied. This
process can be expressed as
\begin{equation}
\rho_\mathrm{out}=\Lambda^U_A(\rho_{AB})\otimes\frac{I_{A'}}{2}.
\end{equation}
where
$\Lambda^U_A(\rho_AB)=\mathrm{Tr}_{A'}(U_{AA'}\rho_{AB}\otimes|0\rangle_{A'}\langle0|U_{AA'}^{\dagger})$.
Therefore, the super-activation of the two channels $\Lambda_1=I$
and $\Lambda_2$ is in fact local creation of quantum correlation by
the channel $\Lambda^U_A$.

We will then focus on situations where no pairwise quantum
correlations are induced between $A$ and $\tilde B$ or between $A'$
and $\tilde B$, where $\tilde B$ is the system on Bob's side. The
mechanism for super-activation of QCP under this condition is
totally different from that for local creation of quantum
correlation. In this case, the two states $\xi_1$ and $\xi_2$ used
for checking nonzero QCP of the channel $\Lambda_1\otimes\Lambda_2$
as in Eq. (\ref{criterion}) should satisfy
\begin{equation}
[\xi_1^A,\xi_2^A]=0,\ [\xi_1^{A'},\xi_2^{A'}]=0.\label{rc}
\end{equation}
Here $\xi_i^{A(A')}=\mathrm{Tr}_{A'(A)}\xi_i$ are reduced density
matrices. In the following, we will see that, even limited to the
situation where no pairwise quantum correlation is induced, the
super-activation of QCP can still be observed for most of the
channels.

We first discuss the situation where $\Lambda_1$ and $\Lambda_2$ are
the qubit channels. According to
\cite{PhysRevLett.107.170502,PhysRevA.85.032102}, $\Lambda$ has
zero-QCP only when it is a completely decohering channel
$\Lambda^{\mathrm{CD}}$ or a unital channel $\Lambda^{\mathrm I}$.
We will focus on the super-activation of QCP for two unital
channels. A unital channel is defined as a channel which keeps the
identity operator invariant
\begin{equation}
\Lambda^{\mathrm I}\equiv\{\Lambda:\Lambda(I)=I\}.
\end{equation}
It has been proved that any unital channel of a qubit is unitarily
equivalent to a Pauli channel \cite{222939961} $\Lambda^\mathrm
I(\cdot)=v\Lambda^{\mathrm{Pauli}}(u^{\dagger}(\cdot)u)v^{\dagger}$.
Here the Kraus operators of a Pauli channel are proportional to
Pauli matrices
\begin{equation}
\Lambda^{\mathrm{Pauli}}(\cdot)=\sum_{i=0}^3\lambda_i\sigma_i(\cdot)\sigma_i,\label{Pauli}
\end{equation}
where $\sigma_0=I$, $\sigma_{1,2,3}$ are the three Paule matrices,
$\lambda_i\geq0$, $\lambda_0\geq\lambda_{1,2,3}$, and
$\sum_{i=0}^3\lambda_i=1$. Therefore, it is adequate to consider the
super-activation of QCP for Pauli channels.

\emph{Theorem 2.} When limited to the case where no pair-wise
quantum correlation is induced, the super-activation of QCP can be
observed for unital qubit channels which are not one of the
following cases: (a) both channels are identical isotropic channels,
and (b) one of the channels is a completely depolarizing channel.

\emph{Proof.} For case (b), consider that channel $\Lambda_2$ is a
completely depolarizing channel. The left hand side of Eq.
(\ref{criterion}) equals to
$[\Lambda_1(\xi_1^A),\Lambda_1(\xi_2^A)]\otimes I^{A'}/2$, which
vanishes under the constraint of Eq. (\ref{rc}) and the
super-activation of QCP can not be observed. In the following, we
focus on the cases that none of the two channels are completely
depolarizing channels. $\xi_1$ and $\xi_2$ in Eq. (\ref{criterion})
is chosen as $\xi_i=|\Phi_i\rangle\langle\Phi_i|$, where
$|\Phi_1\rangle=(|00\rangle+|11\rangle)/\sqrt2$ and
$|\Phi_2\rangle=(|0+\rangle+|1-\rangle)/\sqrt2$. In Pauli
presentation, we have
\begin{eqnarray}
\xi_1&=&\frac14(I\otimes
I+\sigma_1\otimes\sigma_1-\sigma_2\otimes\sigma_2+\sigma_3\otimes\sigma_3)\nonumber\\
\xi_2&=&\frac14(I\otimes
I+\sigma_1\otimes\sigma_3+\sigma_2\otimes\sigma_2+\sigma_3\otimes\sigma_1).\label{test_state}
\end{eqnarray}
For $\Lambda_1$ being a Pauli channel, we have
$\Lambda_1(\sigma_i)=a_i\sigma_i$ with
$a_i=\lambda^{(1)}_0+2\lambda^{(1)}_i-\sum_i\lambda^{(1)}_i$,
$i=1,2,3$. Similarly, $\Lambda_2(\sigma_i)=b_i\sigma_i$.
Consequently, $\Lambda_1\otimes\Lambda_2(\xi_1)=\frac14(I\otimes
I+a_1b_1\sigma_1\otimes\sigma_1-a_2b_2\sigma_2\otimes\sigma_2+a_3b_3\sigma_3\otimes\sigma_3)$
and $ \Lambda\otimes\Lambda(\xi_2)=\frac14(I\otimes
I+a_1b_3\sigma_1\otimes\sigma_3+a_2b_2\sigma_2\otimes\sigma_2+a_3b_1\sigma_3\otimes\sigma_1)$.
Therefore, Eq. (\ref{criterion}) means that
\begin{equation}
a_1a_3(b_1^2-b_3^2)=0,b_1b_3(a_1^2-a_3^2)=0.\label{channel1}
\end{equation}
By alternating the subscript $1,2,3$ in Eq. (\ref{test_state}) and
adopting Eq. (\ref{criterion}), we have
\begin{eqnarray}
a_2a_3(b_2^2-b_3^2)=0,b_2b_3(a_2^2-a_3^2)=0,\nonumber\\
a_1a_2(b_1^2-b_2^2)=0,b_1b_2(a_1^2-a_2^2)=0.\label{channel2}
\end{eqnarray}
It means that the super-activation of QCP for channels violating
Eqs. (\ref{channel1}) or (\ref{channel2}) can be detected by the
states $\xi_1$ and $\xi_2$ in form of Eq. (\ref{test_state}) or
those obtained by alternating the subscript $1,2,3$ in Eq.
(\ref{test_state}). Then we are left with the following situations.

(1) $|a_2|=|a_3|=a>0$, $|b_2|=|b_3|=b>0$, and $a_1=b_1=0$. Noticing
that conditions $a_1=0$ and $\lambda_0\geq\lambda_{1,2,3}$ lead to
$a_{2,3}\geq0$, we have $a_2=a_3=a>0$ and $b_2=b_3=b>0$. $\Lambda_1$
and $\Lambda_2$ are both projecting-and-depolarizing channels
$\Lambda^{\mathrm{Dp}}$, which takes a state
\begin{equation}
\zeta=(I+\vec r\cdot\vec\sigma)/2,\label{Bloch_1_qubit}
\end{equation}
to
$\Lambda^{\mathrm{Dp}}_\chi(\zeta)=[I+\chi(r_2\sigma_2+r_3\sigma_3)]/2$,
where $\chi=a,b$. In other words, the channel projects the Bloch
vector $\vec r$ onto $x-y$ plain and then shorten it,
$\Lambda^{\mathrm{Dp}}_\chi\equiv\Lambda^\mathrm
D_\chi\circ\Lambda^p$.

(2) $|a_2|=|a_3|=a>0$, $b_2=b_3=a_1=0$, and $b_1\neq0$. Similar
discussions as in Case (1) give that $a_2=a_3=a>0$ and $b_1=b>0$.
$\Lambda_1=\Lambda^{\mathrm{Dp}}_a$ while
$\Lambda_2=\Lambda_b^{\mathrm{DPD}}\equiv\Lambda^\mathrm
D_b\circ\Lambda^{\mathrm{PD}}$ is equivalent to a completely
dephasing channel $\Lambda^{\mathrm{PD}}$ followed by a depolarizing
channel $\Lambda^\mathrm D_b$.

(3) $|a_1|=|a_2|=|a_3|>0$ and $|b_1|=|b_2|=|b_3|>0$. If
$a_1=-a_2=a_3\neq0$, we have
$\lambda_0=\lambda_1=\lambda_3>\lambda_2$. This channel is
equivalent to the isotropic channel with
$\lambda_1=\lambda_2=\lambda_3>\lambda_0$ by a unitary operator
$\sigma_2$. Therefore, this case is equivalent to
$a_1=a_2=a_3=a\neq0$ and $b_1=b_2=b_3=b\neq0$. $\Lambda_1$ and
$\Lambda_2$ are isotropic channels.

(4) $b_1=b_2=b_3=0$. Channel $\Lambda_2$ is a completely
depolarizing channel, which we have already considered.

Since two completely decohering channels do not have the property of
super-activation of QCP, let $\Lambda_1$ not be a completely
decohering channel. Cases (1) and (2) include the channels obtained
by alternating the subscripts 1,2,3. In the following, we will
derive the commutative states $\xi_1$ and $\xi_2$ satisfying Eq.
(\ref{rc}) to detect the super-activation for these cases, and prove
that such states does not exist for Case (3) with $a_1=b_1$.

Writing a two-qubit state in form
\begin{equation}
\xi_k=\frac14[I\otimes I+\sum_{i=1}^3r_i^{k}\sigma_i\otimes
I+\sum_{i=1}^3s_i^{k}I\otimes\sigma_i+\sum_{i,j=0}^3T^{k}_{ij}\sigma_i\otimes\sigma_j],\label{pauli2}
\end{equation}
where $k=1,2$, we can present $\xi_k$ as $\xi_k=\{\vec r^{k},\vec
s^{k},\hat T^{k}\}$. It is worth noting that the commutation
$[\xi_1,\xi_2]$ can also be written as the Bloch decomposition
\begin{equation}
[\xi_1,\xi_2]=\frac{\tilde
i}{16}[\sum_{i=1}^3\alpha_i\sigma_i\otimes
I+\sum_{i=1}^3\beta_iI\otimes\sigma_i+\sum_{i,j=0}^3\Gamma_{ij}\sigma_i\otimes\sigma_j],\label{commut}
\end{equation}
where $\tilde i$ is the imaginary unit. By using the commutation for
Pauli matrices, we have $[\sigma_i\otimes
I,\sigma_{i'}\otimes\sigma_{j'}]=[\sigma_i,\sigma_{i'}]\otimes\sigma_{j'}$,
$[\sigma_i\otimes I,\sigma_i'\otimes
I]=[\sigma_i,\sigma_{i'}]\otimes I$, and
$[\sigma_i\otimes\sigma_j,\sigma_{i'}\otimes\sigma_{j'}]=[\sigma_i,\sigma_{i'}]\otimes
I\delta_{jj'}+I\otimes[\sigma_j,\sigma_{j'}]\delta_{ii'}$, where
$\delta_{jj'}$ is the Kroneker delta. Therefore,
\begin{eqnarray}
\Gamma_{ij}&=&(\vec r^1\times\vec T^2_{r_j}-\vec r^2\times\vec
T^1_{r_j})_i+(\vec s^1\times\vec T^2_{s_i}-\vec s^2\times\vec
T^1_{s_i})_j,\nonumber\\
\vec\alpha&=&\vec r^1\times\vec r^2+\sum_j\vec T^1_{r_j}\times\vec
T^2_{r_j},\nonumber\\
\vec\beta&=&\vec s^1\times\vec s^2+\sum_i\vec T^1_{s_i}\times\vec
T^2_{s_i},
\end{eqnarray}
where $\vec T^k_{r_j}=\{T^k_{1j},T^k_{2j},T^k_{3j}\}$ and $\vec
T^k_{s_i}=\{T^k_{i1},T^k_{i2},T^k_{i3}\}$. Then $[\xi_1,\xi_2]=0$ is
equivalent to
\begin{equation}
\vec\alpha=\vec\beta=0,\hat\Gamma=0.\label{detect2}
\end{equation}
Meanwhile, $[\xi_1^A,\xi_2^A]=0$ and $[\xi_1^{A'},\xi_2^{A'}]=0$
give that
\begin{equation}
\vec r^1\times\vec r^2=0,\vec s^1\times\vec s^2=0.\label{detect3}
\end{equation}

For Cases (1) and (2), as well as Case (3) with $a\neq b$, we can
find states $\xi_1$ and $\xi_2$ satisfying Eqs. (\ref{detect2}) and
(\ref{detect3}) such that Eq. (\ref{criterion}) is violated.
Precisely, for Case (1) we chose $\xi_k$ to be
\begin{eqnarray}
\vec r^1&=&\vec s^1=\{0,0,r\},\vec r^2=\vec
 s^2=\{0,0,nr\},\nonumber\\
\hat T^1&=&\hat T^2=\mathrm{diag}\{t,t,t\}.\label{detect}
\end{eqnarray}
Direct calculation leads to
$[\Lambda^{\mathrm{Dp}}_a\otimes\Lambda_b^{\mathrm{Dp}}(\xi_1),\Lambda_a^{\mathrm{Dp}}\otimes\Lambda_b^{\mathrm{Dp}}(\xi_2)]=\tilde
iabrt(1-n)(b\sigma_2\otimes\sigma_1+a\sigma_1\otimes\sigma_2)\neq0$.
For Case (2), we chose $\xi_1$ and $\xi_2$ to be
\begin{eqnarray}
\vec s^1&=&\vec r^2=\{0,r,r\},\vec r^1=\vec s^2=0,\nonumber\\
\hat T^1&=&(\hat T^{2})^{\mathrm
T}=\{0,0,0;t,0,0;-t,0,0\},\label{detect1}
\end{eqnarray}
and then
$[\Lambda^{\mathrm{Dp}}_a\otimes\Lambda_b^{\mathrm{DPD}}(\xi_1),\Lambda_a^{\mathrm{Dp}}\otimes\Lambda_b^{\mathrm{DPD}}(\xi_2)]=2\tilde
iabrt\sigma_1\otimes\sigma_1\neq0$. For Case (3) with $a\neq b$, we
choose $\xi_1$ and $\xi_2$ as in Eq. (\ref{detect1}), and then
$[\Lambda^{\mathrm{Iso}}_a\otimes\Lambda_b^{\mathrm{Iso}}(\xi_1),\Lambda_a^{\mathrm{Iso}}\otimes\Lambda_b^{\mathrm{Iso}}(\xi_2)]=2\tilde
iabrt(a-b)\sigma_1\otimes\sigma_1\neq0$.

Now we only need to prove that for two identical isotropic channels,
the super-activation of QCP can not be detected when under the
constraint that no pairwise quantum correlation is induced. From the
property of the isotropic channels
\begin{equation}
\Lambda^\mathrm{Iso}_a\otimes\Lambda^\mathrm{Iso}_a(\xi_k)=\{a\vec
r^{k},a\vec s^{k},a^2\hat T^{k}\},\label{DB}
\end{equation}
and consequently, the commutation of the output states can be
written as
$[\Lambda^\mathrm{Iso}_a\otimes\Lambda^\mathrm{Iso}_a(\xi_1),
\Lambda^\mathrm{Iso}_a\otimes\Lambda^\mathrm{Iso}_a(\xi_2)]=
\{\vec\alpha^\mathrm{Iso},\vec\beta^\mathrm{Iso},
\hat\Gamma^\mathrm{Iso}\}$, where $\vec\alpha^\mathrm{Iso}=a^2\vec
r^1\times\vec r^2+a^4\sum_j\vec T^1_{r_j}\times\vec T^2_{r_j},\
\vec\beta^\mathrm{Iso}=a^2\vec s^1\times\vec s^2+a^4\sum_i\vec
T^1_{s_i}\times\vec T^2_{s_i}$, and
$\hat\Gamma^\mathrm{Iso}=a^3\hat\Gamma$. Therefore, Eqs.
(\ref{detect2}) and (\ref{detect3}) imply
$\vec\alpha^\mathrm{Iso}=\vec\beta^\mathrm{Iso}=0$ and
$\hat\Gamma^\mathrm{Iso}=0$. It means that for two identical
isotropic channels, the super-activation of QCP can not be detected
under the constraint that no pairwise quantum correlation is
induced. This completes the proof of theorem 2.

It is quite interesting that, a completely dephasing channel can
activate the QCP of a depolarizing channel, even under the
constraint that no pairwise quantum correlation is induced. More
precisely, we consider initial classical-classical state
$\rho=\sum_{i=0}^3|\Phi_i\rangle_{AA'}\langle\Phi_i|\otimes|i\rangle_{\tilde{B}}\langle
i|$, where $|\Phi_0\rangle=(|01\rangle-|10\rangle)/\sqrt2$,
$|\Phi_3\rangle=(|-0\rangle-|+1\rangle)/\sqrt2$, and
$|\Phi_{1,2}\rangle$ are the same as in Eq. (\ref{test_state}). Let
$\Lambda_A$ be a depolarizing channel with $\lambda_0=(1+3a)/4$ and
$\lambda_1=\lambda_2=\lambda_3=(1-a)/4$ in Eq. (\ref{Pauli}), and
$\Lambda_{A'}$ be a completely dephasing channel with
$\lambda_0=\lambda_3=1/2$ and $\lambda_1=\lambda_2=0$. The output
state $\Lambda_A\otimes\Lambda_{A'}\otimes I_{\tilde{B}}(\rho)$ has
positive quantum correlation since
$[\Lambda_A\otimes\Lambda_{A'}(\Phi_1),\Lambda_A\otimes\Lambda_{A'}(\Phi_2)]=-\frac{\tilde
i}{4}a^2\sigma_2\otimes\sigma_0$, which is nonzero for $a\neq0$. Let
us look more closely at the correlation structure of the input state
$\rho$. Clearly, the pairwise quantum correlation between any two
particles is zero. However, the genuine quantum correlation is
nonzero. This is because the entanglement between the qubit $A$ and
the composite system $A'{\tilde{B}}$. Measuring $\tilde{B}$ on basis
$\{|i\rangle\}$ and locally operating $A$ and $A'$ can distill a
singlet of qubits $A$ and $A'$. Therefore, the super-activation of
QCP can be understood as transfering the genuine quantum correlation
to the quantum correlation between the bipartition $AA'$ and
$\tilde{B}$. Notice that the generated quantum correlation is still
genuine correlation, because the pairwise quantum correlation is
still zero in the output state.

Now we have studied the case where $\Lambda_1$ and $\Lambda_2$ are
high-dimension channels. We will briefly show that Theorem 2 does
not hold for the general situation where $\Lambda_1$ and $\Lambda_2$
are qu$d$it channels with $d\geq3$. A qu$d$it channel (with
$d\geq3$) has zero QCP if and only if it is either a completely
decohering channel or an isotropic channel. It is obvious that, when
$\Lambda_2$ is a completely depolarizing channel, Eq.
(\ref{criterion}) holds for any commutative states $\xi_1$ and
$\xi_2$ which satisfy Eq. (\ref{rc}). However, for Case (a) in
Theorem 2 where both $\Lambda_1$ and $\Lambda_2$ are identical
isotropic channels, Eq. (\ref{criterion}) can be violated by some
commutative states $\xi_1$ and $\xi_2$ which satisfy Eq. (\ref{rc}).
An equivalent statement for Theorem 2(a) is that for two commutative
two-qubit states $\xi_1$ and $\xi_2$ which satisfy Eq. (\ref{rc}),
the following equation holds
\begin{equation}
[\xi_1,\xi_2^A+\xi_2^{A'}]=[\xi_2,\xi_1^A+\xi_1^{A'}].\label{commut1}
\end{equation}
The equivalence is obvious when we notice that
$\Lambda^\mathrm{Iso}_a\otimes\Lambda^\mathrm{Iso}_a(\xi_k)=a^2\xi_k+a(1-a)(\xi_k^A+\xi_k^{A'})/2+I^{AA'}/4$.
However, Eq. (\ref{commut1}) does not always hold for $A$ and $A'$
being qu$d$its with $d\geq3$. For example, consider $d=4$ and that
the two commutative two-qu$d$it states are
$\xi_1=[I+T_1(\sigma_1\otimes\sigma_2)_A\otimes(\sigma_3\otimes\sigma_1)_{A'}+T_2(\sigma_3\otimes\sigma_1)_A\otimes(\sigma_3\otimes\sigma_2)_{A'}]/16$
and
$\xi_2=[I+T_1I^A\otimes(\sigma_1\otimes\sigma_2)_{A'}+T_2(\sigma_2\otimes\sigma_3)_A\otimes(\sigma_1\otimes\sigma_1)_{A'}]/16$,
where $T_{1,2}\neq0$. Then Eq. (\ref{commut1}) is violated, since
the left hand side of Eq. (\ref{commut1}) is proportional to $\tilde
iT_1T_2(\sigma_3\otimes\sigma_1)\otimes(\sigma_2\otimes\sigma_0)\neq0$
while the right hand side equals zero. Therefore, Theorem 2(a) does
not hold for channels of higher dimensions. It means that the QCP
high-dimension channels is easier to be super-activated.

\section{Super-additivity of QCP for general channels}

Here we will prove the super-additivity of QCP for general channels.
In the following, we only discuss the problem in the regime that the
quantum correlation $Q$ in Eq. (\ref{QCP}) is quantum discord \cite{JPA34.6899,PhysRevLett.88.017901}, which
is defined as
\begin{equation}
\delta_{B|A}(\rho)=\min_{\{F^A_i\}}S(\rho_{B|\{F^A_i\}})-S_{B|A}(\rho),
\end{equation}
where $S_{B|A}(\rho)=S(\rho)-S(\rho_A)$ with
$S(\rho)=-\mathrm{Tr}(\rho\log_2\rho)$ is conditional entropy,
$\{F_i^A\}$ is a positive operator-valued measure (POVM) on qudit
$A$, $S(\rho_{B|\{F^A_i\}})=\sum_ip_iS(\rho_{B|F^A_i})$ with
$p_i=\mathrm{Tr}(\rho F^{A}_i)$ and
$\rho_{B|F^A_i}=\mathrm{Tr}_A(\rho F^{A}_i)/p_i$ is the average
entropy of $B$ after the measurement.

\emph{Theorem 3.} When two channels $\Lambda_1$ and $\Lambda_2$ are
used paralleled, the QCP of the composite channel
$\Lambda_1\otimes\Lambda_2$ is no less than the sum of the QCP for
the two channels
\begin{equation}
\mathcal Q(\Lambda_1\otimes\Lambda_2)\geq\mathcal
Q(\Lambda_1)+\mathcal Q(\Lambda_2).\label{super_add}
\end{equation}

\emph{Proof.} Let $\rho_1$ and $\rho_2$ be the optimal input state
of $\Lambda_1$ and $\Lambda_2$ respectively, then we have $\mathcal
Q(\Lambda_i)=D(\rho'_i)$ with $\rho'_i\equiv\Lambda_i\otimes
I(\rho_i)$, $i=1,2$. As proved in Ref. \cite{1362905}, the classical
correlation is additive for separable states
$J(\rho\otimes\sigma)=J(\rho)+J(\sigma)$ when $\rho$ is a separable
state. Obviously, $\rho'_i$ are separable states. Therefore, we have
\begin{equation}
\delta(\rho'_1\otimes\rho'_2)=\delta(\rho'_1)+\delta(\rho'_2).\label{add}
\end{equation}
Since $\rho_1\otimes\rho_2$ may not be the optimal input state for
channel $\Lambda_1\otimes\Lambda_2$, from the definition of QCP,
$\mathcal
Q(\Lambda_1\otimes\Lambda_2)\geq\delta(\rho'_1\otimes\rho'_2)=\mathcal
Q(\Lambda_1)+\mathcal Q(\Lambda_2)$. This completes the proof.

From the discussions in the last section, we observe that $\mathcal
Q(\Lambda_1\otimes\Lambda_2)>\mathcal Q(\Lambda_1)+\mathcal
Q(\Lambda_2)$ is quite common. Therefore, we ask the following
question: are there situations where the QCP is additive for
channels with positive QCP? We give a positive answer to this
question by providing a class of channels whose QCP is additive.

Here we define measuring-and-preparing (MP) channel as the operation
which measure on a fixed orthogonal basis and then prepare the qubit
to predefined states conditioned on the measurement results. All MP
channels are unitarily equivalent to
\begin{equation}
\Lambda^{\mathrm{MP}}(\rho)=\sum_{i=0}^{d-1}\langle
i|\rho|i\rangle\eta_i,\label{MP_channel}
\end{equation}
where $\eta_i$ are quantum states. Belonging to the set of MP
channels are the completely decohering channels, as well as the
single-qubit channel with maximum QCP, whose Kraus operators are
$E_0^\mathrm M=|0\rangle\langle0|$ and $E_1^\mathrm
M=|+\rangle\langle1|$. For any MP channels in form of Eq.
(\ref{MP_channel}), the optimal input state to reach the maximum
quantum discord in the output state is
\begin{equation}
\rho^\mathrm{MP}=\sum_{i=0}^{d-1}p_{i}|i\rangle\langle
i|\otimes|i\rangle\langle i|.\label{optimal_input}
\end{equation}
The reason is as follows. Writing the general form of an optimal
input state
$\rho=\sum_{i=0}^{d-1}q_{i}|\phi_i\rangle\langle\phi_i|\otimes|i\rangle\langle
i|$, we have the corresponding output state
\begin{equation}
\rho'=\sum_{i=0}^{d-1}p_{i}\eta_i\otimes\rho_i,\label{output1}
\end{equation}
where $p_i=\sum_jq_j|\langle i|\phi_j\rangle|^2$ and
$\rho_i=\sum_jq_j|\langle i|\phi_j\rangle|^2|j\rangle\langle
j|/p_i$. Eq. (\ref{output1}) can be obtained from
$\Lambda^\mathrm{MP}\otimes I(\rho^\mathrm{MP})$ by local operation
on $B$, which can not increase the quantum discord. Therefore, the
optimal input state should be in form of Eq. (\ref{optimal_input}).

\emph{Theorem 4.} When a MP channel $\Lambda^{\mathrm{MP}}$ and a
completely decohering channel $\Lambda^{\mathrm{CD}}$ are used
paralleled, the QCP of the composed channel equals to that of
$\Lambda^{\mathrm{MP}}$
\begin{equation}
\mathcal
Q(\Lambda^{\mathrm{MP}}\otimes\Lambda^{\mathrm{CD}})=\mathcal
Q(\Lambda^{\mathrm{MP}}).\label{add1}
\end{equation}

\emph{Proof.} Consider the general form of an optimal input state
\begin{equation}
\rho=\sum_{i,j=0}^{d-1}p_{ij}\Pi_{\phi_{ij}}^{AA'}\otimes\Pi_{\psi_{ij}}^{BB'},\label{input}
\end{equation}
where $|\phi_{ij}\rangle=U|ij\rangle$ and
$|\psi_{ij}\rangle=|ij\rangle$. Without loss of generality, we
assume that the basis of the completely decohering channel is
$\{|i\rangle\}$ and that the MP channel is in the form of Eq.
(\ref{MP_channel}). The QCP of the channel $\Lambda^{MP}$ is just
the quantum discord in state $\Lambda^\mathrm{MP}\otimes
I(\rho^\mathrm{MP})$. Notice that
\begin{equation}
\Lambda^{CD}_{A'}(\Pi_{\phi_{ij}}^{AA'})=\sum_{k=0}^{d-1}q_{ij}^k\rho_{ij}^{(k)A}\otimes|k\rangle_{A'}\langle
k|,
\end{equation}
the state $\rho'=\Lambda_{A'}^{CD}(\rho)$ is of form
\begin{equation}
\rho'=\sum_{k=0}^{d-1}|k\rangle_{A'}\langle
k|\otimes\sum_{i,j=0}^1p_{ij}q^k_{ij}\rho_{ij}^{(k)A}\otimes\Pi_{\beta_{ij}}^{BB'}.
\end{equation}
The output state
$\tilde\rho=\Lambda_A^{MP}\otimes\Lambda_{A'}^{CD}(\rho)$ is then
\begin{equation}
\tilde\rho=\sum_{k=0}^{d-1} r_k|k\rangle_{A'}\langle
k|\otimes\rho_k.\label{output}
\end{equation}
where $r_k=\sum_{i,j=0}^{d-1}p_{ij}q_{ij}^k$,
$\rho_k=\sum_l\eta_l^A\otimes\xi_{lk}^{BB'}$ and
$\xi_{lk}^{BB'}=\sum_{i,j=0}^1p_{ij}q^l_{ij}\langle
k|\rho_{ij}^{(l)}|k\rangle\Pi_{\beta_{ij}}^{BB'}/r_k$. Obviously,
$\delta_{BB'|A}(\rho_k)\leq\mathcal Q(\Lambda^{MP})$ for
$k=1,2,\cdots,d-1$. We will prove in the following that for states
in the form of Eq. (\ref{output}), the quantum discord of
$\tilde\rho$ is lower bounded by the weighted average quantum
discord of $\rho_k$
\begin{equation}
\delta_{BB'|AA'}(\tilde\rho)\leq\sum_{k=0}^{d-1}
r_k\delta_{BB'|A}(\rho_k).\label{adds}
\end{equation}
Suppose the optimal POVM for $\rho_k$ are
$\{F_k^{(i)}\}_{i=0}^{N_k-1}$ respectively. By building a POVM on
qubits $A$ and $A'$ as
$\{G^{(l)}\}_{l=0}^{\sum_kN_k-1}=\{|k\rangle\langle k|\otimes
F_k^{(i)}\}_{k=0,\cdots,d-1}^{i=0,\cdots,N_k-1}$, we have
\begin{eqnarray}
S(\tilde\rho_{BB'|\{F^{(l)}_{AA'}\}})&=&\sum_l\tilde p_l
S(\frac{\mathrm{Tr}_{AA'}(G^{(l)}_{AA'}\tilde\rho)}{\tilde
p_l})\nonumber\\
&=&\sum_{ik} r_k \tilde q_{ki}
S(\frac{\mathrm{Tr}_{A'}(F_{k,A'}^{(i)}\rho_k)}{\tilde q_{ki}})\nonumber\\
&=&\sum_{k=0}^{d-1}
r_kS(\rho_{k,BB'|\{F_k^{(i)}\}}).
\end{eqnarray}
Meanwhile, direct calculation leads to
$S(\tilde\rho_{BB'|AA'})=\sum_{k=0}^{d-1} r_kS(\rho_{k,BB'|AA'})$.
Consequently,
$S(\tilde\rho_{BB'|\{G^{(l)}_{AA'}\}})-S(\tilde\rho_{BB'|AA'})=\sum_{k=0}^{d-1}
r_k\delta_{BB'|A}(\rho_k)$. By noticing that $\{G^{(l)}\}$ may not
be the optimal POVM for the quantum discord of $\tilde\rho$, we have
proved Eq. (\ref{adds}). Since Eq. (\ref{input}) is a general form
of optimal input state, the above discussion shows that
\begin{equation}
\mathcal
Q(\Lambda^{\mathrm{MP}}\otimes\Lambda^\mathrm{CD})\leq\mathcal
Q(\Lambda^{\mathrm{MP}}).\label{suadd}
\end{equation}
Combining Eq. (\ref{suadd}) with theorem 3, we finally reach Eq.
(\ref{add1}). This completes the proof of theorem 4.

It should be noticed that both MP channel and CD channel are
coherence-breaking channels. A CD channel takes any state to a state
which is diagonal on a fixed basis, and thus the coherence between
different state basis is broken. For a MP channel, coherence is
broken during the measurement, and so does the genuine quantum
correlation. Even through the preparation process in the MP channel
can rebuild the bipartite quantum correlation, the genuine quantum
correlation, which enable the super-additivity of QCP, can not be
rebuilt. This is the reason why the QCP of a MP channel and a CD
channel is additive. Therefore, we conjecture that the QCP of
channels which are neither MP channels nor CD channels are
super-additive.

\section{Conclusion}

We have investigated the effect of super-additivity of
quantum-correlating power, which means that using two channels
together is more efficient in creating quantum correlations than
using them separately. Two zero-QCP channels have the property of
super-activation of QCP except the trivial cases that the two
channels are both completely deochering channels or unitary
channels. This result shows that super-activation of QCP is a fairly
common effect for local channels. We also prove the super-additivity
of QCP for general local channels and find a class of quantum
channels whose QCP is additive.

Super-additivity of QCP is a collective effect. The genuine quantum
correlation is observed in the initial state which can detect the
super-activation of QCP for a CD channel paralleled with a
depolarizing channel. Meanwhile, the QCP of a MP channel and a CD
channel is additive since both of them are coherence-breaking
channels, which break the genuine quantum correlation. Therefore, we
conjecture that genuine quantum correlation is responsible for such
effect. This provide a new perspective to look at the concept of
genuine quantum correlation, which is still an open problem in
quantum information theory. From this point of view, our study can
shed light on both classification of quantum channels and the
structure of quantum correlation in multipartite states.

\begin{acknowledgments}
This work was supported by the NKBRSFC under grants Nos.
2011CB921502, 2012CB821305, 2009CB930701, 2010CB922904, NSFC under
grants Nos. 10934010, 60978019, 11175248 and NSFC-RGC under grants Nos.
11061160490 and 1386-N-HKU748/10.
\end{acknowledgments}

\bibliography{apssamp}

\end{document}